\documentclass[runningheads]{llncs}
\pdfoutput=1
\usepackage{graphicx}
\usepackage[marginal]{footmisc}
 
\usepackage{makecell}
\usepackage{multirow}
\usepackage{subfigure}
\usepackage{caption}
\begin{document}
\title{Fusing Medical Image Features and Clinical Features with Deep Learning for Computer-Aided Diagnosis}
%
\titlerunning{Fusing Medical Image Features and Clinical Features for CAD}
%
\author{Songxiao Yang\inst{1} \and
Xiabi Liu\inst{1} \and
Zhongshu Zheng\inst{1} \and
Wei Wang\inst{2} \and
Xiaohong Ma\inst{3} 
}
\authorrunning{Yang et al.}
%
\institute{Beijing Institute of Technology, China \\
          \and
          Xuanwu Hospital, Capital Medical University, China \\
          \and
          National Cancer Center/National CLinical Research Center for Cancer/Cancer Hospital, 
          Chinese Academy of Medical Sciences and Peking Union Medical College, China
}
%
\maketitle              
\begin{abstract}
    Current Computer-Aided Diagnosis (CAD) methods mainly depend on medical 
    images. The clinical information, which usually needs to be considered 
    in practical clinical diagnosis, has not been fully employed in CAD. 
    In this paper, we propose a novel deep learning-based method for fusing 
    Magnetic Resonance Imaging (MRI)/Computed Tomography (CT) images and 
    clinical information for diagnostic tasks. Two paths of neural layers 
    are performed to extract image features and clinical features, respectively, 
    and at the same time clinical features are employed as the attention 
    to guide the extraction of image features. Finally, these two modalities 
    of features are concatenated to make decisions. We evaluate the proposed 
    method on its applications to Alzheimer’s disease diagnosis, mild 
    cognitive impairment converter prediction and hepatic microvascular 
    invasion diagnosis. The encouraging experimental results prove the 
    values of the image feature extraction guided by clinical features and 
    the concatenation of two modalities of features for classification, 
    which improve the performance of diagnosis effectively and stably. 
\end{abstract}
\section{Introduction}
In recent years, there has been remarkable progress on Computer-Aided Diagnosis 
(CAD), especially with the help of deep learning technologies \cite{litjens2017survey}. Some studies 
have shown that the inclusion of CAD system in the diagnostic process can 
improve the performance of image diagnosis by reducing inter-observer variation \cite{sahiner2007malignant,singh2011computer}
and providing quantitative support for clinical decisions, such as biopsy 
recommendations \cite{giger2013breast}.

Medical images are really important for diagnosis. But they are not just the 
information that we should consider in the diagnosis. The clinical information 
is also important and necessary in practical clinical diagnosis. So, it is 
better to fuse medical image and clinical information in CAD system. The main 
methods of fusing these two modalities of information in CAD can be divided into 
three categories. The first category is to concatenate original clinical data 
with image features. Pacheco et al. \cite{pacheco2020impact} proposed a method to detect skin cancers 
by concatenating clinical data with image features to improve the accuracy. 
Mobadersany et al. \cite{mobadersany2018predicting} proposed a method to predict the cancer patient’s 
survival, which uses VGG19 network and Cox proportional hazards model to extract 
features from pathological images. Then the genomic biomarkers data and image 
features are combined with a fully connected layer. Zhen et al. \cite{zhen2020deep} used 
convolutional neural networks (CNNs) to classify liver tumors with enhanced MR 
images, unenhanced MR images and clinical data. Liu et al. \cite{liu2020deep} developed a 
deep learning system to provide diagnosis for skin diseases based on 
concatenating skin photographs’ features and associated medical histories. 
Ning et al. \cite{ning2020open} proposed a method to diagnose coronavirus diseases based on 
concatenating CT features, clinical features and laboratory-confirmed severe 
acute respiratory syndrome coronavirus 2 (SARS-CoV-2) clinical status and use 
penalized logistic regression to classify hybrid features. Tognetti et al. \cite{tognetti2020new} 
proposed a deep learning architecture to classify atypical melanocytic skin 
lesions by concatenating dermoscopic image features and clinical data. Yamamoto 
et al. \cite{yamamoto2020deep} proved that the integration of clinical data can improve the 
performance of diagnosis of osteoporosis from hip radiographs. Sandhu et al. \cite{sandhu2020automated} 
proved that concatenating clinical, demographic data and imaging features can 
improve the performance of diagnosis of nonproliferative diabetic retinopathy 
(NPDR). The second category is to transform original clinical data into a matrix 
and combine it with the image. Sharma et al. \cite{sharma2020classification} proposed a preprocessing method 
that transforms 1D clinical data vector to 2D graphical image, which allows 2D 
CNNs to diagnose breast cancers based on the overlapping of transformed clinical 
data and image data. The third category is to employ clinical information and 
medical image in two independent stages. Yin et al. \cite{yin2020auxiliary} proposed a method that 
firstly according to the clinical manifestation of the patient, then the MRI 
images of the patients who could not be given a clear diagnosis were put into a 
convolution neural network (CNN) to be tested and finally conclude. The 
combination of two types of information improved the accuracy of Parkinson’s 
Disease diagnosis by 7$\%$.

The medical images and clinical information are not only complementary but also 
correlated. Some clinical features can be reflected in images, and some others 
cannot be. The existing methods mainly consider the complementary relationship 
between these two types of information. In this paper, we propose a new deep 
neural network to employ both complementary and correlated relationship between 
the medical images and clinical information for improving the accuracy of medical 
diagnosis. It is composed of two correlated paths of neural layers for extracting 
clinical features and image features and a fully connected layer for merging two 
modalities of features to make decisions for diagnosis. In image feature 
extraction, we embed clinical features as the attention to guide the finding of 
more discriminative image features. We evaluate the proposed approach on its 
applications to Alzheimer’s Disease (AD) diagnosis, Mild Cognitive Impairment (MCI) 
converter prediction and hepatic microvascular invasion diagnosis. 

The main contributions of this paper are summarized as follows.
\begin{enumerate}
    \item We proposed a new deep network architecture that fuses medical images 
    and clinical information to perform medical diagnosis, in which clinical 
    features are employed as the attention to guide image feature extraction 
    and two modalities of features are merged for the final diagnosis. The 
    reasonability of our design for this architecture is confirmed by ablation 
    studies.

    \item The proposed deep network for medical diagnosis is applied to AD 
    diagnosis, MCI converter prediction, and hepatic microvascular invasion 
    diagnosis. It yields beneficial results in these three tasks, which 
    proves the effectiveness and the robustness of our method.
\end{enumerate}

\section{The Proposed Approach}
\subsection{The Whole Architecture}
\begin{figure}
    \centering
    \includegraphics[scale=0.33]{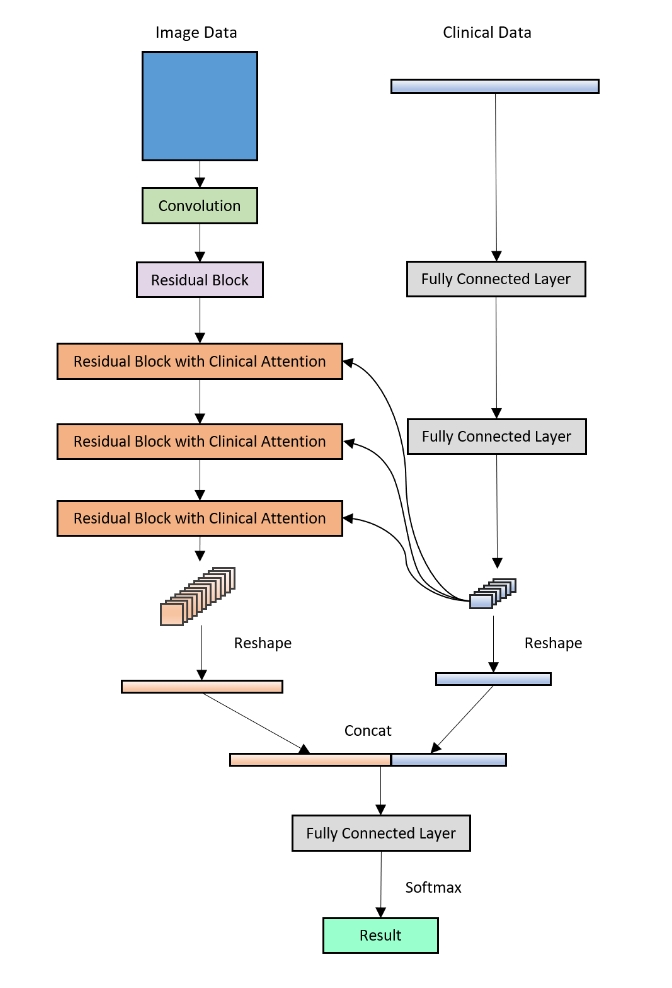}
    \caption{The proposed deep architecture of fusing medical images and clinical information for medical diagnosis} 
    \label{fig1}
\end{figure}

We propose a new deep neural network that fuses medical image and clinical 
information to achieve a more accurate diagnosis. The architecture of our 
deep network is illustrated in Fig. \ref{fig1}, which has two correlated computation 
paths for feature extraction, one for extracting image features and the other 
for extracting clinical feature. As shown in the right part of Fig. \ref{fig1}, we 
construct 2 fully connected layers to extract the features from clinical data. 
These clinical features are inserted into the image feature extraction module 
shown in the left part of Fig. \ref{fig1} as the attention to guide the extraction of 
image features. Considering that clinical features can corelate with different-scale 
image features, we implement residual block with clinical attention for the 
last three blocks of ResNet since the semantic hierarchy of first block is too 
low. Finally, two modalities of features are merged to make decisions for 
diagnosis through a fully connected layer. 

\begin{figure*}
    \centering
    \subfigure[]
    {
    \begin{minipage}[b]{.35\textwidth}
    \flushleft
    \includegraphics[scale=0.18]{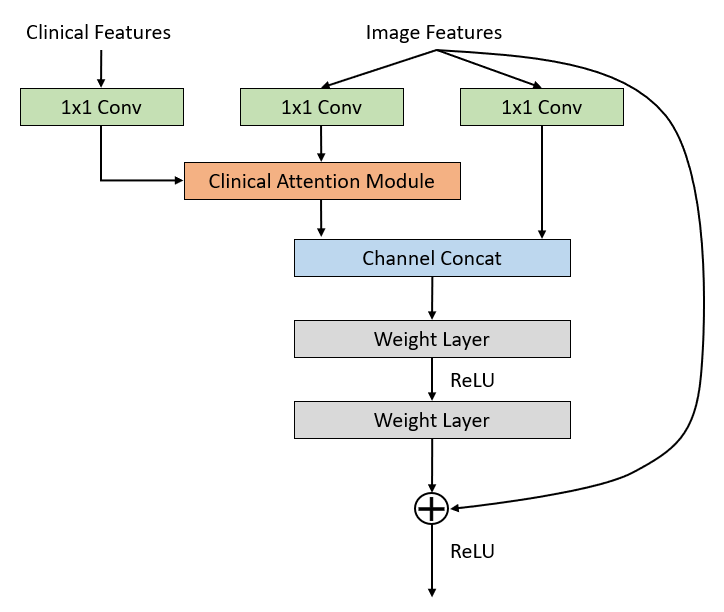}
    \label{fig2a}
    \end{minipage}
    }
    \subfigure[]
    {
    \begin{minipage}[b]{.6\textwidth}
    \flushright
    \includegraphics[width=\textwidth]{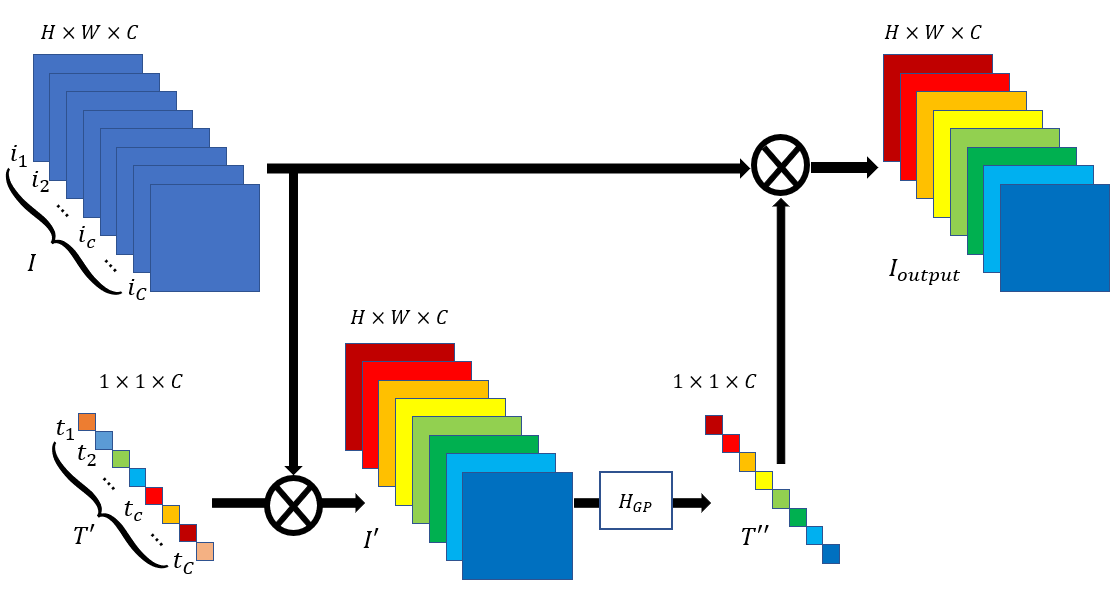}
    \label{fig2b}
    \end{minipage}
    }
    \caption{Residual Block with Attention: (a) the whole structure, (b) the clinical attention module in (a)} 
    \label{fig2}
\end{figure*}

\subsection{Residual Block with Clinical Attention}
As described before, some clinical features can be reflected in images. For 
example, in the diagnosis of Alzheimer’s Disease (AD), the Neuro Filament L 
(NFL) is in direct proportion to the extent of brain shrinkage. In order to 
utilize the correlated clinical features to help finding the attentive features 
in medical images, we insert a clinical feature attention module into the 
residual block of ResNet50 deep network. The 
core part of ResNet50 is the residual block. We modify it to make clinical 
features guide the image feature extraction for improving the accuracy of 
diagnosis. The modified residual block is called Residual Block with Clinical 
Attention (RBCA) and is shown in Fig. \ref{fig2}. 

The inputs of RBCA are medical image features and clinical features. For 
different diagnostic tasks, different clinical features could be reflected in 
different scales of image features. In order to deal with this problem of scale 
selection, we let the original image features concatenated with the image 
features processed by the clinical features and let the neural network learn to 
choose truly useful features from the combination of them. Considering this 
doubles the dimension of features and increase the model complexity greatly, we 
reduce the dimension of inputted image features by a half with $1\times1$ convolution. 
The operations above can keep our representation sparse and make RBCA scale 
insensitive. 

Then the reduced image features are processed by clinical features in Clinical 
Attention Module (CAM), the detail of which is shown in Fig. \ref{fig2b}. At first, to 
align these two modalities of information by channel, we use a convolution layer 
with the kernel size of 1 to process clinical features. Let $T’$ be the processed 
clinical feature vector, $I$ be the inputted image features, $C$ be the number of 
channels of $I$ and $T’$, $H$ and $W$ be the height and the width of image feature map, 
respectively. Each dimension of $I$ is multiplied with correspond element of $T’$ to 
do the similarity measure between each dimension of image features and each 
dimension of clinical features. Let $I’$ be the resultant feature map. To get the 
similarity between each channel of image features and clinical features, we use 
global average pooling to average each resultant feature map. Let $H_{GP} (.)$ be 
the global average pooling function, then we have
\begin{equation}
    H_{GP}(i'_{c})=\frac{1}{H\times W}\sum\limits_{j=1}^{H}\sum\limits_{k=1}^{W}i'_c(j,k),
\end{equation}
where $i’_c(j,k)$ is the value at position $(j,k)$ of $c$-th feature $i’_c$. Such 
channel statistic can be viewed as a collection of the local descriptors, whose 
statistics contribute to express the whole image \cite{hu2018squeeze}. 

Finally, the CAM outputted features and the reduced image features are 
concatenated to pass through two layers of convolution to obtain the final 
result. In this way, the more discriminative features in the image will be 
enhanced and play more influence in the diagnosis.

\subsection{Decision Making Based on Combined Features}
In the last subsection, we employ the clinical features that can be reflected in 
images for improving the extraction of image features. However, there are also 
some useful clinical features that cannot be reflected in the image, such as Tau 
in the diagnosis of AD. It means that we should concatenate the clinical features 
with the image features to do the final classification. In this paper, the merged 
features pass through a fully connected layer and Softmax function to obtain the 
classification possibilities for each category. Then the category with the 
highest probability is taken as the classification result. 

As described in Section 2.1, there are some other works of combining clinical 
data and medical images for diagnosis. However, they usually use the original 
clinical data. The number of attributes in original clinical data is usually 
much smaller than the image features. For example, in our experiments of AD 
diagnosis, original clinical data has only 237 attributes while the dimension 
of image features are 2048. This makes the clinical information drowned in the 
sea of image features and could be neglected in the classification. To solve 
this problem, we extract high dimensional deep features from clinical attributes. 
The number of extracted clinical features is close to that of image features.

\subsection{Learning Algorithm}
As we are dealing with classification problem, we choose the popular loss 
function of categorical Cross-Entropy (CE) to perform learning. Based on the 
CE loss, we apply the Back Propagation (BP) algorithm with stochastic gradient 
descent to update the parameters of our model.

\section{Experiments}
We conduct three groups of experiments on three diagnostic tasks to demonstrate 
the effectiveness of our proposed approach. In the training of our model, we 
use He initialization \cite{he2015delving} and Adam optimization and set the learning rate as 
0.0001 and the number of training epochs as 100. The algorithm is implemented 
in PyTorch framework and runs over a GPU server with Nvidia RTX 2080Ti. 

In the experiments, we conduct ablation studies to demonstrate the reasonability 
of the design of our approach. Our main contributions on the deep architecture 
for AD diagnosis exist in two aspects, guiding the image feature extraction by 
employing clinical features as the attention and concatenating two modalities 
of features in the final classification. In order to confirm whether these two 
new operations are really effective, we perform diagnosis by using each of four 
following options of our deep architecture and compare their performance:
\begin{enumerate}
    \item Only Image: only image features are employed
    \item Image + Clinical: In the classification, image features and 
    clinical features are used, but image feature extraction is not guided by 
    clinical features
    \item Full Model: our full model including two new operations
\end{enumerate}
Furthermore, we conduct 5-fold cross validation on these three tasks. 
\subsection{AD Diagnosis}
AD is characterized as a genetically complex and irreversible neurodegenerative 
dis-order and often found in persons aged over 65. AD versus Normal Cognitive 
(NC) is a typical problem within AD diagnosis.
We evaluate our proposed method for AD diagnosis on AD Neuroimaging Initiative 
dataset (ADNI 1 1.5T)\footnote{Data used in preparation of this article were obtained from the Alzheimer’s Disease Neuroimaging Initiative (ADNI) database (adni.loni.usc.edu). As such, the investigators within the ADNI contributed to the design and implementation of ADNI and/or provided data but did not participate in analysis or writing of this report. A complete listing of ADNI investigators can be found at: \url{http://adni.loni.usc.edu/wp-content/uploads/how\_to\_apply/ADNI\_Acknowledgement\_List.pdf}}. 
From ADNI 1 1.5T, we construct a 2D MRI dataset with 5640 slices, 4530 of them 
are used for training and 1110 of them for testing. The patients in the training 
set and the test set have no intersection. Inspired by \cite{zhang2015detection}, we selected 10 
slices from the coronal position of every MRI image, with y-value range 
between 130 and 140, because these slices are corresponding with informative 
part of brain for AD diagnosis. We resize each slice to $256\times256$. As for the 
clinical data, we choose a part of the biospecimen data by a senior doctor 
with 14 years’ experiences in AD diagnosis. Total 237 clinical data are 
obtained, which are listed in supplement material. 

The results of AD diagnosis are reported in Table \ref{ADNI1}. We can see that the clinical 
data is very valuable for AD diagnosis. Introducing clinical features lead to 
better performance on all the criterions, even they are only concatenated with 
image features. However, the images also contain many useful information for 
diagnosis, which cannot be reflected in clinical features. By appending our 
attention from clinical information and combining two modalities of features 
to utilize some clinical information that are necessary for diagnosis but 
cannot be reflected in images, the classification accuracy based on only images 
can be improved greatly, as shown in the comparisons between the results in the 
last two rows in Table \ref{ADNI1}.

\begin{table}
    \centering
    \caption{The results of ADNI1 dataset}
    \label{ADNI1}
    \begin{tabular}{|c|c|c|c|c|c|c|}
        \hline
        Task & Approach & ACC & Sensitivity & Specificity & PPV & NPV \\
        \hline
        \multirow{3}*{AD vs. NC} & Image Only & 66.73\% & 54.29\% & 73.68\% & 53.05\% & 74.64\% \\
        \cline{2-7} 
        ~ & Image+Clinical & 92.70\% & 86.18\% & 95.74\% & 90.35\% & 93.74\% \\
        \cline{2-7} 
        ~ & Full Model & \textbf{95.52\%} & \textbf{92.41\%} & \textbf{96.96\%} & \textbf{93.25\%} & \textbf{96.53\%}\\  
        \hline  
    \end{tabular}
\end{table}

\subsection{MCI Converter Prediction}
Recently, an increasing number of studies on AD research begin to address 
classification of MCI to AD conversion (MCI-AD) and MCI non-conversion (MCI-NC) 
patients based on the high-resolution brain imaging data.
We also choose a part of MRI images of MCI patients from ADNI2, with a dataset of 
800 slices, 640 of them are used for training and 160 of them for testing. These 
images are used for the classification of MCI versus MCI-AD converter. The image 
data selection method and preprocessing are the same with ADNI1. For ADNI2, total 
234 clinical data are obtained, which are also listed in supplement material. 
The results of MCI converter prediction are reported in Table \ref{ADNI2}. We can observe 
the same phenomenon as those in the first experiment, i.e. concatenating clinical 
features with image features can improve the accuracy of MCI converter prediction, 
and using clinical features as the attention can extract better image features 
and further improve the accuracy of prediction. 

\begin{table}
    \centering
    \caption{The results of ADNI2 dataset}
    \label{ADNI2}
    \begin{tabular}{|c|c|c|c|c|c|c|}
        \hline
        Task & Approach & ACC & Sensitivity & Specificity & PPV & NPV \\
        \hline
        \multirow{3}*{\makecell*[c]{MCI vs. MCI-AD\\converter}} & Image Only & 67.41\% & 42.92\% & 82.33\% & 79.25\% & 71.25\% \\
        \cline{2-7} 
        ~ & Image+Clinical & 70.63\% & 50.42\% & 83\% & 75.67\% & 73.67\% \\
        \cline{2-7} 
        ~ & Full Model & \textbf{74.06\%} & \textbf{55.73\%} & \textbf{85.13\%} & \textbf{78.06\%} & \textbf{77.38\%}\\  
        \hline  
    \end{tabular}
\end{table}

\subsection{Hepatic Microvascular Invasion}
Liver cancer is the second leading cause of cancer-related deaths worldwide and 
hepatocellular carcinoma (HCC) represents the most common primary liver 
cancer \cite{el2007hepatocellular,wang2016global}, hence the diagnosis of hepatic microvascular invasion is quite 
an important task.
We used 1394 2D liver CT images from 139 patients of hepatic microvascular 
invasion to conduct our third group of experiments. Total 29 preoperative clinical data are obtained, 
which are also listed in supplement material. This dataset is used for the 
diagnosis of hepatic microvascular invasion (HMI). The results of hepatic microvascular invasion diagnosis 
are shown in Table \ref{HMI}. Again, concatenating image and clinical features improve 
the accuracy of hepatic microvascular invasion diagnosis and introducing 
clinical attention module into image feature extraction can further improve the 
performance of the hepatic microvascular invasion diagnosis.

\begin{table}
    \centering
    \caption{The results of HMI dataset}
    \label{HMI}
    \begin{tabular}{|c|c|c|c|c|c|c|}
        \hline
        Task & Approach & ACC & Sensitivity & Specificity & PPV & NPV \\
        \hline
        \multirow{3}*{\makecell*[c]{Hepatic Microv-\\ascular Invasion\\ Diagnosis}} & Image Only & 71.18\% & 88.32\% & 34.76\% & 74.23\% & 48.57\% \\
        \cline{2-7} 
        ~ & Image+Clinical & 75.47\% & \textbf{90.09\%} & 46.12\% & 77.24\% & 69.11\% \\
        \cline{2-7} 
        ~ & Full Model & \textbf{79.17\%} & 88.84\% & \textbf{58.91\%} & \textbf{81.76\%} & \textbf{78.89\%}\\  
        \hline  
    \end{tabular}
\end{table}

\section{Conclusions}
This paper has proposed a deep network architecture to fuse medical image and 
clinical information for Alzheimer’s Disease (AD) diagnosis, MCI versus 
MCI-converter prediction and liver tumor benign/malignant diagnosis, in which 
clinical features are used as the attention for image feature extraction 
and combined with the image features to perform classification. To our best 
knowledge, this is the first work of utilizing clinical data as the attention 
by deep learning to guide the image feature extraction in computer-aided 
diagnosis. Our approach yields a good performance on the ADNI1 dataset, 
ADNI2 dataset and our private dataset. The ablation studies confirm that 
1) the clinical information is very valuable and necessary for medical 
diagnosis; 2) the better image features can be obtained under the guidance 
of clinical features, and 3) two modalities of features are complementary 
and should be combined for improving the accuracy of diagnosis.

%
%
%
\bibliographystyle{splncs04}
\bibliography{mybibliography}
%




\end{document}